\documentclass[aps,prx,twocolumn,groupedaddress,showkeys]{revtex4}
\usepackage{amsmath}
\usepackage{amssymb}
\usepackage{graphicx}
\usepackage{xcolor}

\begin{document}

\title{Hidden Anisotropy in the Drude Conductivity of 
Charge Carriers\\ with Dirac-Schr\"odinger Dynamics}

\author{Maxim Trushin,$^a$ Antonio H. Castro Neto$^a$, Giovanni Vignale$^{a,b}$, and Dimitrie Culcer$^c$}
\affiliation{$^a$Centre for Advanced 2D Materials, National University of Singapore, 6 Science Drive 2, Singapore 117546}
\affiliation{$^b$Department of Physics and Astronomy, University of Missouri, Columbia, Missouri 65211}
\affiliation{$^c$School of Physics and Australian Research Council Centre of Excellence in Low-Energy Electronics Technologies, The University of New South Wales, Sydney 2052, Australia}

\date{\today}

\begin{abstract}
We show that the conductivity of a two-dimensional electron gas can be {  intrinsically} anisotropic despite isotropic Fermi surface, energy dispersion, and disorder configuration. 
{  In the model we study, the anisotropy stems from the interplay between Dirac and Schr\"odinger features combined in a special two-band Hamiltonian describing the quasiparticles similar to the low-energy excitations in phosphorene.}
As a result, even scalar isotropic disorder scattering alters the nature of the carriers and results in anisotropic transport. Solving the Boltzmann equation exactly for such carriers with point-like random impurities we find a hidden knob to control the anisotropy just by tuning either the Fermi energy or temperature. 
Our results are expected to be generally applicable beyond the model studied here, {  and should stimulate further search for the alternative ways to control electron transport in advanced materials.}

\end{abstract}

\keywords{anisotropic conductivity, Boltzmann equation, pseudospin, phosphorene}
\maketitle

{\em Introduction. ---}
The electronic properties of two-dimensional (2D) materials \cite{bhimanapati2015recent}
 and heterostructures \cite{novoselov20162d} inevitably fascinate condensed matter theorists \cite{neto2009electronic,sarma2011electronic} offering
a few already solved as well as still puzzling problems including but not limited to 
the conductivity minimum in graphene \cite{adam2007self,trushin2010finite,dean2010boron}, 
superconductivity in twisted double-layer graphene \cite{cao2018unconventional},
edge-state conductivity \cite{kane2005z,hasan2010colloquium} in 2D topological insulators \cite{moore2010birth} 
leading to the quantum spin and anomalous Hall effects \cite{shen2017}, the superconducting proximity effect \cite{fu2008superconducting}, the valley-Hall effect in dichalcogenides \cite{mak2014valley}, negative magneto-resistance \cite{dai2017negative, breunig2017gigantic}, and, most recently, 
unconventional second-order electrical response \cite{yasuda2016, he2018}. 
Conventional models for conductivity of a 2D electron gas 
rely on the well established concepts of group velocity, Fermi surface, electronic density of states (DOS), disorder scattering \cite{ando1982electronic} and Berry curvature \cite{sundaram1999wavepacket}.
However, a complete understanding of electron transport in various 
2D materials often requires details of the effective Hamiltonian describing
the electron motion through the crystal lattice.
The most famous example is monolayer graphene, where
both electrons and holes have vanishing effective mass
mimicking 2D massless Dirac fermions \cite{neto2009electronic}.
Other known 2D materials, where charge carriers are described by a non-trivial two-band Hamiltonian, 
include bilayer graphene \cite{mccann2006landau},
hexagonal 2D boron nitride \cite{ribeiro2011stability},
monolayer group-VI dichalcogenides \cite{xiao2012coupled},
and, most recently, phosphorene \cite{PRB2015pereira}.
These effective Hamiltonians often possess an additional degree of freedom with non-trivial texture
(e.g. pseudospin \cite{PRL2011trushin}, due to e.g. inequivalent sublattices, valleys, or angular momentum), 
and, therefore, may offer a hidden knob to control electron transport
leaving all other conventional parameters, like the Fermi surface or disorder, unchanged.

{  Our work has been inspired by phosphorene} --- a 2D layer of black phosphorus --- a material expected to have
a great potential in optoelectronics because of its high electron mobility and optical absorption \cite{ACSNano2014liu,ACR2016lu,NatRevMat2016carvalho}.
The effective two-band Hamiltonian for carriers in phosphorene 
\cite{PRB2015pereira,PRL2014rodin,Fukuoka2015} has an intriguing property:
in the lowest order in the two-component momentum $(k_x,k_y$) the Hamiltonian
has a Dirac-like structure for one component (say, $k_y$) but
Schr\"odinger-like structure for another ($k_x$) \cite{ezawa2015highly}.
There should therefore be a qualitative difference in scattering of electrons moving in $x$ and in $y$ directions. 
In effect scattering alters the nature of the charge carriers, since a change in wave vector can transform the particle
from Dirac-like to Schr\"odinger-like and vice versa. Such a change can be brought about even by an isotropic scalar potential. 
This feature sets our system apart from e.g. graphene, which has both a sublattice and a valley pseudospin, yet neither of these are associated with anisotropy in the conductivity.
The effect is not related to an effective mass anisotropy and it was not isolated until now
despite the Boltzmann transport theory developed recently 
for carriers in phosphorene \cite{zare2017thermoelectric}.
We show that such peculiarity in electron scattering can make the electrical conductivity 
anisotropic despite isotropic effective mass and scalar disorder.
Having discovered this effect we call it \textit{hidden anisotropy}.
{  Unfortunately, the hidden anisotropy is overwhelmed by the conventional one in pristine phosphorene because the electron and hole effective masses are strongly anisotropic there.
However, the latter anisotropy can be eliminated by a proper choice of tight-binding parameters in the phosphorene-like lattice model, see  Appendix  for details.}

\begin{figure}
\includegraphics[width=\columnwidth]{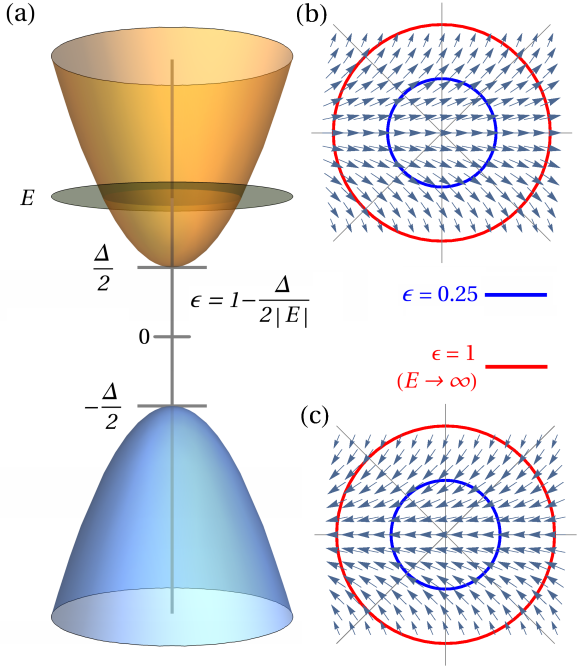}
\caption{\label{fig0} 
(a) The band structure given by the eigenvalues of (\ref{H}) is parabolic and isotropic.
The energy $E$ is measured from the middle of the bandgap and
characterized by the dimensionless parameter $\epsilon$ changing between $0$ (band edge)
and $1$ (theoretical limit $|E| \gg \Delta/2$).
The in-plane pseudospin texture for conduction (b) and valence (c) bands 
suggests anisotropic scattering even for isotropic delta-correlated disorder.
At $\epsilon \ll 1$, it emulates the pseudospin texture for carriers in phosphorene.}
\end{figure}

\begin{figure}
\includegraphics[width=\columnwidth]{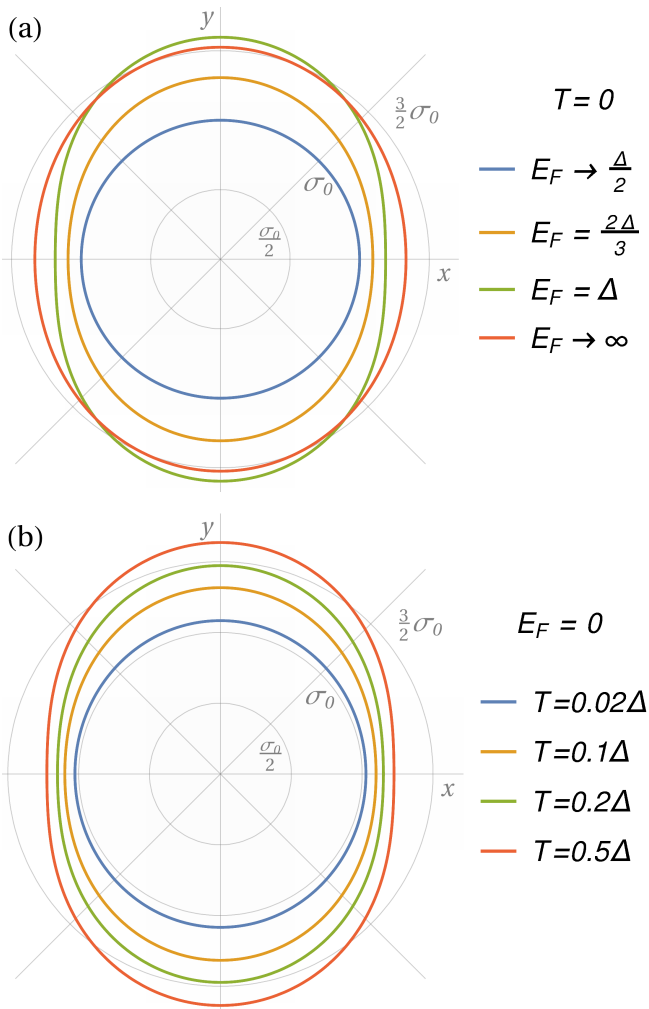}
\caption{\label{fig1} 
Conductivity $\sigma$ in units of {  the conventional Drude conductivity} $\sigma_0=e^2 \tau_0 n/m $
as a function of the current direction
in the metallic (a) and semiconducting (b) regimes. {  The Fermi energy is counted from the middle of the bandgap,
i.e. it intersects the bottom of the conduction band at  $E_F=\Delta/2$.}
The limit $E_F\to\infty$ is nonphysical and shown for the sake of completeness.}
\end{figure}

{\em Model. ---} Having phosphorene \cite{PRB2014rudenko}
and phosphorene-like materials \cite{carvalho2015} in mind we,
however, address a much more general problem.
Assume we have a 2D electron gas confined
in a 2D conductor whose lattice symmetry determines an effective two-band Hamiltonian
that combines Dirac- and Schr\"odinger-like features given by
\begin{equation}
\label{H}
 \hat H=\left(
\begin{array}{cc}
\frac{\hbar^2 k_y^2}{2m}  &  \frac{\Delta_{k_x}}{2} -i \hbar k_y \sqrt{\frac{\Delta_{k_x}}{2m}} \\
 \frac{\Delta_{k_x}}{2} + i \hbar k_y \sqrt{\frac{\Delta_{k_x}}{2m}} &
 -\frac{\hbar^2 k_y^2}{2m}
\end{array} \right),
\end{equation}
where $m$ is the effective mass, and $\Delta_{k_x}=\Delta + \hbar^2k_x^2/m$ with
$\Delta$ being the fundamental bandgap. The spectrum has two branches 
$E^{\pm}_k=\pm E$, see Fig. \ref{fig0}(a), with $E= \frac{\Delta}{2} + \frac{\hbar^2 k^2}{2m}$ and $k^2=k_x^2+k_y^2$,
corresponding to two eigenstates
\begin{equation}
\label{psi}
 \psi^+_k=\left(\begin{array}{c}
\cos\frac{\theta}{2}\\
\sin\frac{\theta}{2} e^{i\gamma}
\end{array} \right),\quad
 \psi^-_k=\left(\begin{array}{c}
\sin\frac{\theta}{2}\\
-\cos\frac{\theta}{2} e^{i\gamma}
\end{array} \right).
\end{equation}
The following angular variables have been defined:
\begin{eqnarray}
\label{theta}
 && \cos\theta=\epsilon\sin^2\phi, \quad \cos\gamma= \frac{1-\cos\theta}{\sin\theta},\\
\label{gamma}
 && \sin\gamma=\frac{\sin\phi}{\sin\theta}\sqrt{2\epsilon(1-\cos\theta)},\quad \tan\phi= \frac{k_y}{k_x},
\end{eqnarray}
and $\epsilon=1-\frac{\Delta}{2|E|}$, $0\leq \epsilon \leq 1$.
To make the hidden anisotropy apparent we cast our Hamiltonian into the form
$\hat H=\sum_i\hat\sigma_i {\cal H}_i$, where $\hat\sigma_i$ are the Pauli matrices
representing the pseudospin, $i=\{x,y,z\}$,
and ${\cal H}_i$ are the pseudomagnetic field components.
The pseudospin eigenstate expectation values then read
\begin{eqnarray}
\label{sigmaxz}
 && \langle\hat\sigma_x\rangle = \pm \left(1-\epsilon \sin^2 \phi\right), \quad \langle\hat\sigma_z\rangle = \pm\epsilon \sin^2\phi,\\
\label{sigmay}
 && \langle\hat\sigma_y\rangle = \pm\sin\phi \sqrt{2\epsilon(1-\epsilon \sin^2\phi)},
\end{eqnarray}
and the corresponding pseudospin textures are shown Fig. \ref{fig0}(b,c).
The direction of a pseudospin vector strongly depends on $\epsilon$ at $\phi\neq 0,\pi$,
whereas at $\phi=0,\pi$ the pseudospin keeps its orientation along the $x$-axis no matter how large $\epsilon$ is.
Hence, the pseudospin texture forms a preferred direction along which
the texture remains collinear and has no influence on scattering. This direction corresponds to the $x$-axis in our case,
see also Appendix for comparison with the pseudospin texture on a phosphorene-like lattice.
If electrons are moving in any other direction, then the pseudospin texture reduces electron backscattering
that facilitates the transport and makes conductivity anisotropic, see Fig. \ref{fig1}.

{\em Boltzmann equation.---}
We write the Boltzmann equation for electrons assuming that the electron
steady-state distribution function, $f_k$, is independent of the spatial coordinates
and can be written as a sum of equilibrium and non-equilibrium 
terms, $f_{E\phi}=f^0_E+f^{x,y}_{E\phi}$, where $f^{x,y}_{E\phi} \ll f^0_E$.
The latter inequality is justified for weak electric field ${\cal E}_{x,y}$ (linear response regime).
The Fermi-Dirac distribution $f^0_E$ is characterized by the Fermi energy $E_F$
and temperature $T$ (in energy units).
The Boltzmann equation has two components written as
\begin{eqnarray}
\label{to_solve}
&& e{\cal E}_{x,y}\left\{\begin{array}{c}
\cos\phi\\
\sin\phi \end{array} \right\} \sqrt{\frac{2E}{m}\epsilon}
 \frac{\partial f^0_E}{\partial E} \\
&&
\nonumber =\int\limits_0^{2\pi}d\phi'
 \int\limits_{\Delta/2}^\infty 
 \frac{d E' m}{(2\pi\hbar)^2} w_{EE'\phi\phi'}
 \left(f^{x,y}_{E'\phi'}-f^{x,y}_{E\phi}\right),
\end{eqnarray}
where $w_{EE'\phi\phi'}$ 
is the golden-rule transition probability given by
$  w_{EE'\phi\phi'} =\frac{2\pi}{\hbar}n_i |U_{EE'\phi\phi'}|^2 \delta\left(E -  E' \right)$.
Here, $n_i$ is the impurity concentration, and 
$U_{EE'\phi\phi'}$ is the 
scattering potential matrix element. The simplest case is that of a delta-shaped scattering potential 
with constant Fourier transform $u_0$ (see Appendix).
Equation (\ref{to_solve}) can be solved 
using the {\em Ansatz} depending on
the electric field direction
\begin{equation}
\label{solution}
 f^{x,y}_{E\phi}
 = e{\cal E}_{x,y}\tau_{xx,yy}
\left\{\begin{array}{c}
\cos\phi\\
\sin\phi \end{array} \right\} \sqrt{\frac{2E}{m}\epsilon}
\left(-\frac{\partial f^0_E}{\partial E}\right),
\end{equation}
where the momentum relaxation times are given by
\begin{subequations}
\begin{align}
\label{tauxx} \tau_{xx}(\phi)= & \frac{\tau_0}{1-\epsilon\left(\frac{1}{4} +\frac{1-\epsilon}{2}
 \sin^2 \phi \right)}, \\
\label{tauyy}
\tau_{yy}(\phi) = & \tau_0\frac{1+a_\epsilon \sqrt{1-\epsilon\sin^2\phi}}
 {1-\epsilon\left(\frac{1}{4} +\frac{1-\epsilon}{2}\sin^2 \phi \right)},
\end{align} 
\end{subequations}
and $\tau_0^{-1}=n_i m u_0^2/\hbar^3$. 
The relaxation times depend on the electric field direction
despite isotropic Fermi surface and disorder.
The parameter that makes $\tau_{xx}$ and $\tau_{yy}$
qualitatively different is $a_\epsilon$, which we refer to as the "anomaly parameter''. Its derivation can be found in Appendix.
The anomaly parameter is a non-monotonic function of $\epsilon$ but 
in the low-energy limit it allows for linear approximation
$a_\epsilon=\epsilon/2$, $\epsilon \ll 1$.
Note that our solution (\ref{solution}) holds for arbitrary $a_\epsilon$.  The momentum relaxation time equals the Bloch lifetime (obtained from Fermi's golden rule) along $x$, along which the motion is Schr\"odinger-like, but they are different along all other directions.

{\em Electrical conductivity. ---}
We calculate the electrical current density as
\begin{equation}
 j_{x,y} = 2e
 \int\limits_0^{2\pi}d\phi
 \int\limits_{\Delta/2}^\infty 
 \frac{d E m}{(2\pi\hbar)^2} f^{x,y}_{E\phi}
 \left\{\begin{array}{c}
\cos\phi\\
\sin\phi \end{array} \right\} \sqrt{\frac{2E}{m}\epsilon},
\end{equation}
where the factor $2$ is due to the spin degeneracy,
and $f^{x,y}_{E\phi}$ is given by (\ref{solution}).
The conductivity tensor has two components $\sigma_{xx}$ and $\sigma_{yy}$ given by
\begin{subequations}
	\begin{align}
 \label{sigmaxx}
 \sigma_{xx} = & \frac{e^2 \tau_0}{\pi\hbar^2} \int\limits_{\Delta/2}^\infty dE
 \left(E-\frac{\Delta}{2}\right) \left(-\frac{\partial f^0_E}{\partial E}\right) I_{xx}(E), \\
 \label{sigmayy}
 \sigma_{yy} = &\frac{e^2 \tau_0}{\pi\hbar^2} \int\limits_{\Delta/2}^\infty dE
 \left(E-\frac{\Delta}{2}\right) \left(-\frac{\partial f^0_E}{\partial E}\right) \\
\nonumber & \times \left[I_{yy}(E)+a_\epsilon I_{yy}^a(E)\right].
	\end{align}
\end{subequations}
Here, $\phi$ was integrated out, and resulting expressions $I_{xx}(E)$, $I_{yy}(E)$,
and $I_{yy}^a(E)$ are given in Appendix.

We consider two limiting transport regimes: metallic ($E_F\geq \Delta/2$, $T\to 0$)
and semiconducting ($E_F=0$, i.e. the semiconductor is intrinsic).
In both cases, the conductivity is anisotropic, and the anisotropy increases
with electron energy that can be controlled by either Fermi energy
or temperature, depending on the transport regime. 
The conductivity scaled by $\sigma_0=e^2 n \tau_0/m$ is plotted in Fig. \ref{fig1}.
{  Here, $\sigma_0$ is the conventional Drude conductivity for Schr\"odinger carriers, and $n$ is the electron concentration.}  
In the metallic regime Eqs. (\ref{sigmaxx}--\ref{sigmayy}) can be simplified
using the substitution $\left(-\partial f^0_E/\partial E\right) \to \delta\left(E -E_F \right)$, so that 
$\sigma_{xx}= \sigma_0 I_{xx}(E_F)$, and $\sigma_{yy}= \sigma_0 \left( I_{yy} + a_\epsilon I_{yy}^a \right)_{E=E_F}$.
Further simplifications can be performed assuming low doping and using the Taylor-expansion in terms
of $1-\Delta/(2E_F)$ (that is $\epsilon$ taken at $E=E_F$). The anisotropy can be then characterized by the ratio
\begin{equation}
\frac{\sigma_{xx}}{\sigma_{yy}}\to \frac{1+3\Delta/(2E_F)}{4}, \quad 1-\frac{\Delta}{2E_F} \ll 1.
\end{equation}
At $E_F = \Delta/2$ the anisotropy disappears but the 
conductivity vanishes itself in this limit. Hence, the conductivity is anisotropic as long as it is non-zero.
The anisotropy is therefore an {\em  intrinsic} property of our model.

{\em Discussion. ---}
The pseudospin texture depicted in Fig. \ref{fig0} determines the conductivity anisotropy shown in Fig. \ref{fig1}.
Thanks to the texture, disorder affects electrons moving in different directions in a different way.
The scattering probability is reduced when the pseudospin has to change its orientation upon scattering.
Obviously, the electrons moving along $x$-direction experience no probability reduction for backscattering,
as the pseudospin orientation remains the same no matter how large the momentum is.
This maximizes resistance and reduces the conductivity $\sigma_{xx}$ to a minimum.
In contrast, the carriers moving in any direction other than along $x$-axis have to change 
the pseudospin orientation upon backscattering. In particular, the pseudospin texture becomes strongly non-collinear
for carriers moving along $y$-axis so that the back scattering probability is reduced resulting in an inequality $\sigma_{yy}>\sigma_{xx}$. Certainly, the conductivity is determined not only by the backscattering probability,
as the full integral over all possible scattering angles contributes to the electrical resistance. 
Hence, both $\sigma_{xx}$ and $\sigma_{yy}$ increase with energy
as compared with the Drude value $\sigma_0$ because the overall non-collinearity
of the pseudospin texture becomes stronger.
The energy of carriers contributing to the conductance is determined by
either the Fermi energy (in the metallic regime) or by temperature (in the semiconducting regime).
Thus, the anisotropy can be controlled externally by means of doping and/or heating, as demonstrated in Fig. \ref{fig1}.

The pseudospin eigenstate expectation values (\ref{sigmaxz},\ref{sigmay}) taken at $\epsilon \ll 1$
emulate the pseudospin texture for carriers in phosphorene \cite{ezawa2015highly}, see Appendix for references.
In this limit, $\langle\hat\sigma_x\rangle \simeq \pm 1$,
$\langle\hat\sigma_z\rangle \simeq 0$, and $\langle\hat\sigma_y\rangle \simeq \pm\sin\phi \sqrt{2\epsilon}$,
i.e. the in-plane pseudospin collinearity is perfect along the $x$-axis but diminished for other directions,
and the out-of-plane component vanishes as long as the higher-order terms in $\epsilon$ are neglected.
Similar to phosphorene, our Hamiltonian $\hat H$ is time-reversal invariant. Moreover, the eigenstates (\ref{theta},\ref{gamma})
do not produce any Berry curvature or Berry phase. 
This peculiar property of our Hamiltonian can be understood in terms of 
the pseudospin field components ${\cal H}_i$ defined below Eqs. (\ref{theta},\ref{gamma}).
The pseudomagnetic field following the pseudospin direction in Fig. \ref{fig0}(b,c) on a closed trajectory 
in momentum space does not subtend a solid angle and, hence, results in zero Berry phase.
The vanishing Berry curvature is explicitly calculated in Appendix. 
Note the special relation between ${\cal H}_i$ given by $\sqrt{2 {\cal H}_x {\cal H}_z}={\cal H}_y$,
that makes it possible to exclude ${\cal H}_z$ from the Hamiltonian and 
write all the resulting formulas in terms of ${\cal H}_y$ and ${\cal H}_x$ only.
This also determines a special ``equi-pseudospin'' curves in momentum space
along which the pseudospin does not change. 
Obviously, any additional term ${\cal H}'_z \hat \sigma_z$ would drastically change 
the geometry and result in non-zero Berry curvature. 

The velocity operators derived from (\ref{H}), given by
\begin{subequations}
	\begin{align}
\label{vx}
 \hat v_x = & \cos\phi\sqrt{\frac{2E}{m}\epsilon} \, \left(\hat\sigma_x + \frac{1}{2}\tan\gamma \, \hat\sigma_y\right), \\
\label{vy}
 \hat v_y = & \sin\phi\sqrt{\frac{2E}{m}\epsilon} \, \left(\hat\sigma_z + \cot\gamma \, \hat\sigma_y\right),
 	\end{align}
\end{subequations}
resemble the velocity behavior for carriers in phosphorene.
At $k\to0$ (i.e. $\epsilon\to 0$) we find $\tan\gamma\to 0$ and $\hat v_x\to 0$
(Schr\"odinger-like behavior) but $\cot\gamma\to\infty$ and
$\hat v_y \to \hat\sigma_y \sqrt{\frac{\Delta}{2m}}$ (Dirac-like behavior).
This also explains why the conductivity is always highest along the $y$-axis:
The carriers behave like Dirac particles along this direction, which implies a strong reduction in backscattering.
The corresponding group velocities, i.e. the diagonal elements of 
Eqs. (\ref{vx}) and (\ref{vy}) written in the basis of the Hamiltonian eigenstates,
are the same as for a conventional electron gas and given by
$v_x^{\pm}=\pm |v| \cos\phi $, $v_y^{\pm}=\pm |v| \sin\phi$ with $|v|=\sqrt{\frac{2E}{m}\epsilon}$.
Despite the complexity of the internal structure of our Hamiltonian resembling phosphorene,
the spectrum remains isotropic and parabolic. 
This makes it possible to reveal the hidden conductivity anisotropy at least theoretically.
To reveal the effect experimentally in true phosphorene, the band structure anisotropy should be reduced
by applying strain \cite{PRL2014rodin,wang2015electro} or chemical doping \cite{kim2015observation,guo2016tuning},
see also Appendix.

Anisotropic eigenfunctions (\ref{theta},\ref{gamma}) are not something new for
Hamiltonians describing 2D electrons with spin-orbit 
coupling of Rashba \cite{JETPLett1984} and Dresselhaus \cite{PR1955dresselhaus}
types. However, the spin-orbit splitting even being strongly anisotropic by itself
does not lead to anisotropic conductivity \cite{PRB2007trushin}.
To make the conductivity anisotropic in such heterostructures 
one has to break time-reversal invariance by adding, 
e.g. magnetized impurities \cite{PRB2009trushin}.
Our Hamiltonian, in contrast, does not break time-reversal and even does not depend on true spin orientation
but still leads to a strongly anisotropic conductivity. {  Furthermore, whereas crystal Hamiltonians are determined from atomic orbitals using symmetry considerations, they may have special properties when the coupling constants have certain values, as is exemplified by the spin helix state in semiconductors with equal Rashba and Dresselhaus interactions. We have studied the special case of a generic Hamiltonian with parameters tuned to achieve an isotropic energy dispersion, which we regard as a dynamical symmetry.}


We note that Hamiltonian (\ref{H}) must be considered within the quasiclassical 
approximation, where $k_{x,y}$ are just numbers.
Otherwise, we have to deal with the pseudo-differential operators 
that might be a challenging task in the context of scattering.
Moreover, we must assume $\sqrt{k_x^2}=|k_x|$
in the theoretical limit $k \to \infty$ ($\epsilon\to 1$) to keep topology
the same for any energy and avoid breaking time-reversal invariance in our Hamiltonian.
After all, we have to keep the conductivity continuous at $E_F\to \infty$.
This limit is anyway unrealistic and unphysical leading to a model effect seen in Fig. \ref{fig1}(a). 
The conductivity and its anisotropy starts to decrease again with $E_F$ if the latter is too high.
This makes the conductivity non-monotonic as a function of $E_F$, see Appendix.
In the limit of $\Delta/(2E_F) \to 0$ (infinite doping or zero bandgap)
$a_\epsilon=16/(15\pi)$, $I_{xx}=I_{yy}=4/3$, and the anisotropy is
$\sigma_{xx}/\sigma_{yy}\to 1/(1+\frac{64}{45\pi^2})\sim 0.87$, see Appendix.
Formally speaking, the out-of-plane pseudospin components become important at such energies
and reduce the non-collinearity of the pseudospin texture. This regime is not related to phosphorene.

{\em Conclusion.---}
The common belief is that the conductivity anisotropy occurs thanks to either anisotropic Fermi surface or
non-scalar disorder (or both). We have found an interesting example in which anisotropy can be induced solely by the internal structure of the effective Hamiltonian comprising Schr\"odinger and Dirac features. This internal structure does not influence the bands, which always remain parabolic and isotropic, but creates a peculiar pseudospin texture. The texture provides the wave functions with an additional phase depending on the direction of motion. One might think that it is the Berry's phase that is responsible for this effect but the phase is in fact zero. The origin of conductivity anisotropy is therefore hidden in the Hamiltonian much deeper than just the Berry curvature, and becomes apparent by altering between Dirac and Schr\"odinger dynamics due to scattering on disorder. This is the reason why this effect is dubbed ``hidden anisotropy''. This hidden anisotropy can easily be tuned by changing either the Fermi level or the temperature providing a ``hidden knob'' for electron transport control.

{\em Acknowledgements.}
M.T. conceived the project, devised the model, and wrote the draft.
A.H.C.N. provided connection to phosphorene. G.V. and D.C. made connection to the vanishing Berry's phase.
M.T. and A.H.C.N. acknowledge funding support from the Singapore National Research Foundation (NRF).
In particular, M.T. has been supported by the Director's Senior Research Fellowship 
from the CA2DM at NUS (NRF Medium Sized Centre Programme R-723-000-001-281) and by the Gordon Godfrey Bequest while at UNSW.
DC is supported by the Australian Research Council Centre of Excellence in Future Low-Energy Electronics Technologies
(project number CE170100039) funded by the Australian Government.

\appendix

\begin{widetext}

\section{Construction of a general model with isotropic dispersion and anisotropic conductivity}

We look for a model Hamiltonian of the form
\begin{equation}
\hat H = h_x\hat\sigma_x+h_y\hat\sigma_y+h_z\hat\sigma_z\,,
\end{equation}
where $h_x$, $h_y$, $h_z$ are functions of $k_x$ and $k_y$, and $\hat\sigma_{x,y,z}$ are the Pauli matrices.
The dispersion is
\begin{equation}
E_\pm=\pm\sqrt{h_x^2+h_y^2+h_z^2}\,.
\end{equation}

We want the dispersion to be isotropic, i.e., a function of $k^2=k_x^2+k_y^2$.
Furthermore, we want the Hamiltonian to have a special direction in $k$-space along which the pseudospin orientation 
(the expectation value of the vector operator $\hat\sigma$ in the eigenstate basis of $\hat H$) remains constant. 
(The Berry curvature for such a model will vanish.)
This is necessary to make the conductivity anisotropic.

A general way to achieve both requirements is to choose any of the three components of the pseudomagnetic field
$\mathbf{h}$, say $h_y$, to be the geometric average of the other two:
\begin{equation}
h_y=\sqrt{2 h_xh_z}\,.
\end{equation}
With this choice the dispersion takes the form
\begin{equation}
E_\pm=\pm \vert h_x+h_z\vert\,.
\end{equation}
Furthermore, the Hamiltonian takes the form
\begin{equation}
\hat H = \sqrt{2h_xh_y}\left(\hat\sigma_y+\sqrt{\frac{h_x}{2h_z}}\hat\sigma_x+\sqrt{\frac{h_z}{2h_x}}\hat\sigma_z\right)\,,
\end{equation}
from which it is evident that the orientation of the pseudomagnetic field depends only on the ratio $\frac{h_z}{h_x} $ and remains constant on the curves along which $h_z=\beta h_x$, where $\beta=\mathrm{const}$.

Restricting ourselves to polynomials of second order in $k_x$ and $k_y$ the simplest choice for $h_x$ and $h_z$ that guarantees an isotropic dispersion is
\begin{equation}
h_x= a k_x^2 +bk_y^2+ \Delta_x\,, \quad h_z= a k_y^2 +bk_x^2+ \Delta_z\,,
\end{equation}
where $a$, $b$, $\Delta_x$, and $\Delta_y$ are constants.  This gives the isotropic dispersion
\begin{equation}
E_\pm=\pm \vert (a+b)k^2+(\Delta_x+\Delta_z)\vert\,.
\end{equation}
Notice that, while the dispersion depends only on the combinations $a+b$ and $\Delta_x+\Delta_z$, the curves of constant pseudomagnetic field direction depend separately on each of these parameters.

The Hamiltonian discussed in our paper corresponds to the choice
\begin{equation}
a=\frac{\hbar^2}{2m}, \quad \Delta_x = \frac{\Delta}{2}, \quad b=0,\quad \Delta_z=0\,.
\end{equation}
Thus we have
\begin{equation}
h_x = \frac{\hbar^2 k_x^2}{2m}+\frac{\Delta}{2}, \quad h_y=\frac{\hbar k_y}{\sqrt{m}}\sqrt{\frac{\hbar^2 k_x^2}{2m}+\frac{\Delta}{2}}, \quad h_z = \frac{\hbar^2k_y^2}{2m}.
\end{equation}
In the low-energy limit 
\begin{equation}
\frac{\hbar^2}{2m}\left(k_x^2+k_y^2 \right)  \ll \frac{\Delta}{2}
\end{equation}
the pseudomagnetic field preferably polarizes pseudospin along $x$-direction,
and the resulting pseudospin texture becomes similar to the one of phosphorene, see the next section.

\section{Pseudospin texture in true phosphorene}

The low-$k$ expansion of the tight-binding Hamiltonian near
$\Gamma$-point in the first Brillouin zone for carriers 
in phosphorene results in \cite{PRB2015pereira} 
\begin{equation}
 \hat H_0=\left(
\begin{array}{lr}
u_0 + \eta_x k_x^2 + \eta_y k_y^2  & \delta + \gamma_x k_x^2 + \gamma_y k_y^2 + i \chi k_y \\
\delta + \gamma_x k_x^2 + \gamma_y k_y^2 - i \chi k_y & u_0 + \eta_x k_x^2 + \eta_y k_y^2
\end{array} \right),
\end{equation}
where $u_0=-0.42\,\mathrm{eV}$, $\eta_x=0.58\,\mathrm{eV}\,\mathring{\mathrm{A}}^2$,
$\eta_y=1.01\,\mathrm{eV}\,\mathring{\mathrm{A}}^2$, $\delta=0.76\, \mathrm{eV}$,
$\chi=5.25\,\mathrm{eV}\,\mathring{\mathrm{A}}$, $\gamma_x = 3.93\,\mathrm{eV}\,\mathring{\mathrm{A}}^2$,
$\gamma_y=3.83\,\mathrm{eV}\,\mathring{\mathrm{A}}^2$.
The single-particle spectrum is given by 
$E_\pm =u_0 + \eta_x k_x^2 + \eta_y k_y^2 \pm \sqrt{(\delta + \gamma_x k_x^2 + \gamma_y k_y^2)^2 + \chi^2 k_y^2}$,
where $\pm$ stands for conduction (valence) band.
The band gap is then given by $\Delta=2\delta\simeq 1.5$ eV.
The eigenstates read $\psi_\pm=\frac{1}{\sqrt{2}}(1, \pm e^{i\zeta})^T$,
where
\begin{equation}
 \tan \zeta = \frac{\chi k_y}{\delta + \gamma_x k_x^2 + \gamma_y k_y^2}.
\end{equation}
The carriers in phosphorene obviously obey an anisotropic dispersion
and can be described in terms of the corresponding effective masses
\cite{PRB2015pereira} with the ratios given by
\begin{equation}
 \label{masses}
 \frac{m_y^e}{m_x^e} = \frac{\eta_x+\gamma_x}{\eta_y+\gamma_y+\chi^2/\Delta}
 \simeq 0.2;\qquad
 \frac{m_y^h}{m_x^h} = \frac{\gamma_x-\eta_x}{\gamma_y+\chi^2/\Delta -\eta_y}
 \simeq 0.16
\end{equation}
for electrons and holes respectively.

However, the dispersion alone does not match 
the {\em nature} of carriers which is different in different directions.
Indeed, the velocity operators $v_{x,y}$ in the lowest order in $k_{x,y}$ have
the form
\begin{equation}
 \label{vxy}
 \hat v_x=\left(
\begin{array}{lr}
\eta_x   & \gamma_x\\
\gamma_x  &  \eta_x
\end{array} \right)\frac{2k_x}{\hbar},\qquad
 \hat v_y=\left(
\begin{array}{lr}
0   & i \chi\\
-i \chi  &  0
\end{array} \right)\frac{1}{\hbar}.
\end{equation}
Near the band edge, $v_x$ vanishes, whereas $v_y$ remains constant.
Hence, it should be a qualitative difference in carrier motion
along $x$ and $y$ directions.
It is especially apparent if we plot the pseudospin texture in
momentum space by calculating the pseudospin eigenstate values as 
$\langle\psi_\pm \lvert \hat\sigma_x \rvert \psi_\pm \rangle = \pm\cos\zeta$,
$ \langle\psi_\pm \lvert \hat\sigma_y \rvert \psi_\pm \rangle = \pm\sin\zeta$,
where $\hat\sigma_{x,y}$ are the Pauli matrices, and 
$\psi_\pm$ are the eigenstates of $H_0$.
The pseudospin texture is plotted in figure \ref{SIfig1}.

Unfortunately, the conductivity anisotropy associated with the pseudospin texture
is masked by the band anisotropy in phosphorene. Indeed, the ratio $\sigma_{xx}/\sigma_{yy}=m_y/m_x$ within the Drude model
suggests the anisotropy 1:5, whereas the pseudospin texture can provide 
the conductivity anisotropy up to about 3:4.
It is however possible to tune the hoping parameters in the phosphorene-like lattice model
to make the conduction and valence bands isotropic but retain the original pseudospin texture.
We should first neglect $\eta_{x,y}$, as $\eta_{x,y} \ll \gamma_{x,y}$ anyway.
Then, we should satisfy the following equation
\begin{equation}
 \label{makeiso}
 \Delta (\gamma_x  - \gamma_y) = \chi^2
\end{equation}
to make $m_y^{e,h}= m_x^{e,h}$. Note that $\gamma_y$ depends on the hopping parameters $t_1$, $t_2$, $t_3$, and $t_5$,
whereas $\gamma_x$ depends on $t_1$ and $t_3$ only, see Refs. \cite{PRB2015pereira,PRB2014rudenko}.
Hence, changing $t_2$ or $t_5$ we can tune the difference $\gamma_x  - \gamma_y$ to satisfy equation (\ref{makeiso})
and in that way create a lattice model with isotropic effective masses for electrons and holes.
Alternatively, we can utilize substitutional doping \cite{guo2016tuning} to control the conventional anisotropy.

\begin{figure}
\includegraphics[width=0.6\columnwidth]{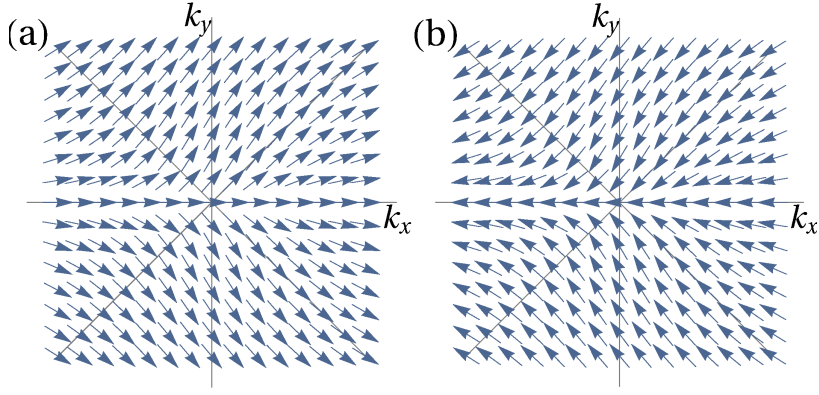}
\caption{\label{SIfig1} 
Pseudospin texture in true phosphorene: (a) conduction band,
(b) valence band. The region in $k$-space shown has dimensions
$1\,\mathring{\mathrm{A}}^{-1}\times 1\,\mathring{\mathrm{A}}^{-1}$.
Our Hamiltonian (1) emulates this texture
near the band edges ($\epsilon \ll 1$).}
\end{figure}

\section{Solution of the Boltzmann equation}

Here we show that our {\em Ansatz} (8) solves equation (7). To begin with, the matrix element for a delta-shaped scattering potential is given by:
\begin{equation}
\label{Uk}
|U_{EE'\phi\phi'}|^2 =
\frac{u_0^2}{2}\left[1+\cos\theta\cos\theta' +
\sin\theta\sin\theta' \cos\left(\gamma-\gamma'\right)\right].
\end{equation}
We first transform equation (\ref{Uk}) using equations (3,4)
as
\begin{eqnarray}
\nonumber && |U_{EE'\phi\phi'}|^2 =
 u_0^2\left[1-\frac{1}{2}\left(\cos\theta+\cos\theta'\right) +\cos\theta\cos\theta'
 \right. \\
&& \left.+\epsilon \sin\phi\sin\phi' \sqrt{1-\cos\theta}\sqrt{1-\cos\theta'}\right] \\
\nonumber && = u_0^2\left[1-\frac{\epsilon}{2}\left(\sin^2\phi+\sin^2\phi'\right) 
+\epsilon^2\sin^2\phi \sin^2\phi'\right.\\
&& \left. +\epsilon\sin\phi\sin\phi' \sqrt{1-\epsilon\sin^2\phi'}\sqrt{1-\epsilon\sin^2\phi}\right].
\label{Uk2}
\end{eqnarray}
Equation (\ref{Uk2}) suggests that $|U_{EE'0\pi}|^2=u_0^2$ for $\phi=0$ and $\phi'=\pi$,
i.e. backscattering is very efficient along $x$-axis.
In contrast, $|U_{EE'\frac{\pi}{2}\frac{3\pi}{2}}|^2=u_0^2(1-2\epsilon+2\epsilon^2)$
for $\phi=\pi/2$ and $\phi'=3\pi/2$, i.e. the backscattering probability is strongly reduced along $y$-axis,
and the strongest reduction occurs at $\epsilon=1/2$. 
Fig. \ref{SIfig2} illustrates the angular dependence of $|U_{EE'\phi\phi'}|^2$ for 
$\phi'=0$, $\phi'=\pi/4$ and $\phi'=\pi/2$.
The dependence is especially interesting at $\phi'=\pi/4$ in the limit $\epsilon=1$, when
it transforms into a strongly asymmetric cardioid-shaped pattern.

If electric field is along $x$-axis, then the Boltzmann equation reads
\begin{equation}
\label{to_solve_x}
 e{\cal E}_x \cos\phi \sqrt{\frac{2E}{m}\epsilon}
 \frac{\partial f^0_E}{\partial E}
 =\int\limits_0^{2\pi}d\phi'
 \int\limits_{\Delta/2}^\infty 
 \frac{d E' m}{(2\pi\hbar)^2} w_{EE'\phi\phi'}
 \left(f^{x}_{E'\phi'}-f^{x}_{E\phi}\right).
\end{equation}
Making use of the delta-function in the scattering probability
$w_{EE'\phi\phi'}$ we have
\begin{equation}
\label{compare}
 -\cos\phi 
 =\int\limits_0^{2\pi}
 \frac{d\phi' m}{(2\pi\hbar)^2} w_{EE'\phi\phi'}
 \left[\cos\phi' \tau_{xx}(\phi') - \cos\phi \tau_{xx}(\phi)\right],
\end{equation}
The first term on the r.h.s. vanishes after integration over $\phi'$.
The second term contains the following integral
\begin{eqnarray}
 \nonumber && \int\limits_0^{2\pi}\frac{d\phi'}{2\pi}
 \left[1-\frac{\epsilon}{2}\left(\sin^2\phi+\sin^2\phi'\right) 
+\epsilon^2\sin^2\phi \sin^2\phi'\right.\\
\nonumber && \left. 
+\epsilon\sin\phi\sin\phi' \sqrt{1-\epsilon\sin^2\phi'}\sqrt{1-\epsilon\sin^2\phi}\right]\\
&& = 1-\frac{\epsilon}{4}-\frac{1-\epsilon}{2}\epsilon\sin^2\phi.
\label{int0}
\end{eqnarray}
Introducing $\tau_0^{-1}=n_i m u_0^2/\hbar^3$ we find 
$\tau_{xx}$ is given by equation (9a), i.e. our solution is correct.

\begin{figure}
\includegraphics[width=\columnwidth]{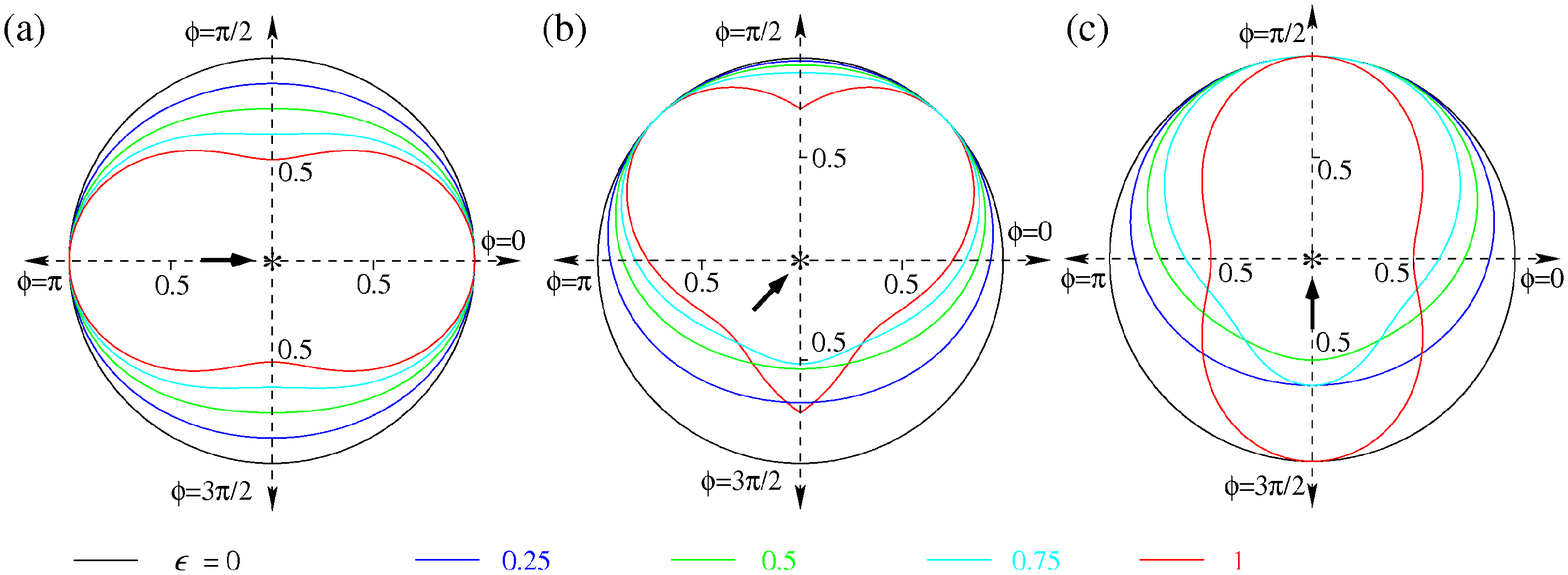}
\caption{\label{SIfig2} 
Scattering matrix element squared (\ref{Uk2}) in units of $u_0^2$
as a function of scattering angle $\phi$
for different incident angles shown by thick arrows:
(a) $\phi'=0$, (b) $\phi'=\pi/4$, and (c) $\phi'=\pi/2$.}
\end{figure}

If electric field is along $y$-axis, then the Boltzmann equation reads
\begin{equation}
\label{to_solve_y}
 e{\cal E}_y \sin\phi \sqrt{\frac{2E}{m}\epsilon}
 \frac{\partial f^0_E}{\partial E}
 =\int\limits_0^{2\pi}d\phi'
 \int\limits_{\Delta/2}^\infty 
 \frac{d E' m}{(2\pi\hbar)^2} w_{EE'\phi\phi'}
 \left(f^{y}_{E'\phi'}-f^{y}_{E\phi}\right),
\end{equation}
that can again be simplified utilizing the delta-function in the scattering probability as
\begin{equation}
\label{compare2}
 -\sin\phi 
 =\int\limits_0^{2\pi}
 \frac{d\phi' m}{(2\pi\hbar)^2} w_{EE'\phi\phi'}
 \left[\sin\phi' \tau_{yy}(\phi') - \sin\phi \tau_{yy}(\phi)\right],
\end{equation}
Using equations (9b) and (\ref{int0}) the r.h.s. of equation (\ref{compare2}) can be rewritten as
\begin{equation}
\label{compare3}
 -\sin\phi 
 =\sin\phi \left[\sqrt{1-\epsilon\sin^2\phi}\left(I_1 + a_\epsilon I_2\right) 
 - \left(1+a_\epsilon\sqrt{1-\epsilon\sin^2\phi}\right)\right],
\end{equation}
where
\begin{equation}
 \label{I1}
 I_1= \epsilon\int\limits_0^{2\pi}\frac{d\phi'}{2\pi}
 \frac{\sin^2\phi' \sqrt{1-\epsilon\sin^2\phi'}}{1-\epsilon
 \left(\frac{1}{4} +\frac{1-\epsilon}{2}\sin^2\phi'\right)},
\end{equation}
and
\begin{equation}
 \label{I2}
 I_2= \epsilon\int\limits_0^{2\pi}\frac{d\phi'}{2\pi}
 \frac{\sin^2\phi' \left[1-\epsilon\sin^2\phi'\right]}{1-\epsilon
 \left(\frac{1}{4} +\frac{1-\epsilon}{2} \sin^2\phi'\right)}.
\end{equation}
Equation (\ref{compare3}) is algebraic with respect to $a_\epsilon$ and the latter
can be determined in terms of $I_1$ and $I_2$ as
\begin{equation}
 \label{ag}
 a_\epsilon=\frac{I_1}{1-I_2}.
\end{equation}
Hence, we have solved the Boltzmann equation using our {\em Ansatz}
with $a_\epsilon$ given by equation (\ref{ag}) and shown in Figure \ref{SIfig3}.

\begin{figure}
\includegraphics[width=0.5\columnwidth]{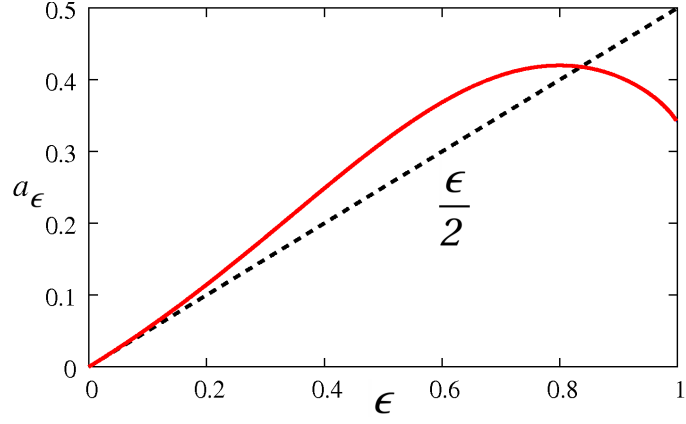}
\caption{\label{SIfig3} 
The anomaly parameter $a_\epsilon$  (red solid curve)
and its low-energy approximation (dashed line). }
\end{figure}

\section{Conductivity integrals}

The conductivity is calculated straightforwardly in terms of the following integrals
\begin{eqnarray}
\nonumber I_{xx} &=& \int\limits_0^{2\pi}\frac{d\phi'}{\pi}
 \frac{\cos^2\phi}{1-\epsilon
 \left(\frac{1}{4} +\frac{1-\epsilon}{2} \sin^2\phi'\right)}\\
 &=&\frac{8}{4-\epsilon+\sqrt{(\epsilon-4)[4+\epsilon(2\epsilon-3)]}},
 \label{Ixx}
\end{eqnarray}
\begin{eqnarray}
\nonumber I_{yy} &=& \int\limits_0^{2\pi}\frac{d\phi'}{\pi}
 \frac{\sin^2\phi}{1 -\epsilon
 \left(\frac{1}{4} +\frac{1-\epsilon}{2} \sin^2\phi'\right)}\\
 &=&\frac{8}{4-\epsilon(2\epsilon-3)+\sqrt{(\epsilon-4)[4+\epsilon(2\epsilon-3)]}},
 \label{Iyy}
\end{eqnarray}
and 
\begin{eqnarray}
\nonumber I_{yy}^a &=& \int\limits_0^{2\pi}\frac{d\phi'}{\pi}
 \frac{\sin^2\phi \sqrt{1-\epsilon\sin^2\phi}}{1-\epsilon
 \left(\frac{1}{4} +\frac{1-\epsilon}{2} \sin^2\phi'\right)}
\end{eqnarray}
with an obvious relation
\begin{equation}
 I_1= \frac{\epsilon}{2}I_{yy}^a.
\end{equation}
The low-temperature conductivity ratio characterizing the anisotropy is given by
\begin{equation}
 \frac{\sigma_{xx}}{\sigma_{yy}} = \left. \frac{I_{xx}}{I_{yy} + a_\epsilon I_{yy}^a }\right|_{E=E_F}
\end{equation}
and plotted in figure \ref{SIfig4}.

\begin{figure}
\includegraphics[width=0.5\columnwidth]{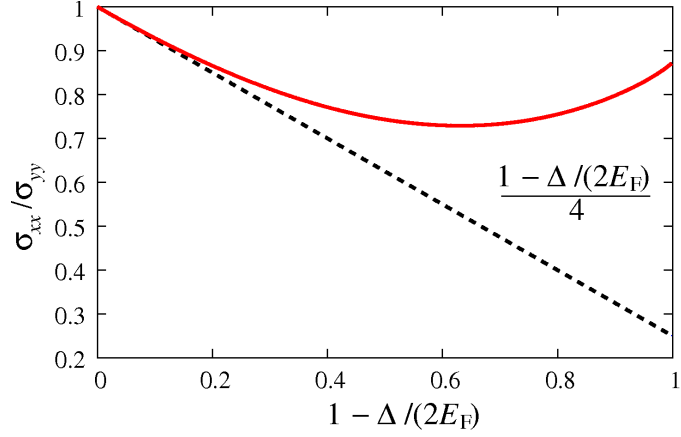}
\caption{\label{SIfig4} 
 Conductivity ratio $\sigma_{xx}/\sigma_{yy}$ (red solid curve)
and its linear approximation (dashed line) in the metallic regime
($E_F>\Delta/2$, $T=0$). Note the non-monotonic dependence. }
\end{figure}

The conductivity can be expressed in units of the Drude conductivity  $\sigma_0=e^2 \tau_0 n/m$,
where the electron concentration is given by
\begin{equation}
n=   \left\{\begin{array}{ll}
\frac{m}{\pi\hbar^2}\left(E_F - \frac{\Delta}{2}\right), & \mathrm{metallic}; \\
\frac{m}{\pi\hbar^2}\left[T\ln \left(1+e^{\frac{\Delta}{2T}}\right) - \frac{\Delta}{2}\right], & \mathrm{semiconducting}.
\end{array} \right.
\end{equation}

\section{Berry curvature calculation}

The Berry curvature vanishes for Hamiltonian (1). Here, we show that
it is indeed so by direct calculation.
In general, Berry curvature is a vector given by a cross-product of two vectors
$\vec\Omega = i \langle\nabla_\mathbf{k}\psi_k\rvert \times  \lvert \nabla_\mathbf{k}\psi_k\rvert \rangle$.
In our case, it has only $z$-component given by
\begin{equation}
 \Omega=-\frac{1}{2}\sin\theta \left(\partial_{k_x}\theta \partial_{k_y}\gamma - \partial_{k_x}\gamma \partial_{k_y}\theta\right),
\end{equation}
where 
\begin{equation}
 \tan\gamma = \sqrt{\frac{2\hbar^2}{m\Delta_{k_x}}}k_y, \quad
 \tan\theta = \frac{\sqrt{\frac{\Delta_{k_x}^2}{4} + \frac{\hbar^2k_y^2}{2m}\Delta_{k_x}}}{\frac{\hbar^2k_y^2}{2m}},\quad
 \Delta_{k_x} = \Delta + \frac{\hbar^2k_x^2}{m}.
\end{equation}
Taking derivatives we have
\begin{equation}
 \frac{\partial \gamma}{\partial k_x} = -\frac{\hbar^2k_xk_y}{m\Delta_{k_x}}\frac{\partial \gamma}{\partial k_y},\quad
 \frac{\partial \gamma}{\partial k_y} = \frac{\sqrt{\frac{2\hbar^2}{m\Delta_{k_x}}}}{1+\frac{2\hbar^2k_y^2}{m\Delta_{k_x}}},
\end{equation}
\begin{equation}
 \frac{\partial \theta}{\partial k_x} = \frac{E}{\sqrt{\frac{\Delta_{k_x}^2}{4} + \frac{\hbar^2k_y^2}{2m}\Delta_{k_x}}}
 \frac{1}{1+\frac{\frac{\Delta_{k_x}^2}{4} + \frac{\hbar^2k_y^2}{2m}\Delta_{k_x}}{\left(\frac{\hbar^2k_y^2}{2m}\right)^2}}
 \frac{\frac{\hbar^2 k_x}{m}}{\frac{\hbar^2k_y^2}{2m}},
\end{equation}
\begin{equation}
 \frac{\partial \theta}{\partial k_y} = -\frac{E}{\sqrt{\frac{\Delta_{k_x}^2}{4} + \frac{\hbar^2k_y^2}{2m}\Delta_{k_x}}}
 \frac{1}{1+\frac{\frac{\Delta_{k_x}^2}{4} + \frac{\hbar^2k_y^2}{2m}\Delta_{k_x}}{\left(\frac{\hbar^2k_y^2}{2m}\right)^2}}
 \frac{\frac{\Delta_{k_x}}{2}\frac{\hbar^2 k_y}{m}}{\left(\frac{\hbar^2k_y^2}{2m}\right)^2}.
\end{equation}
Finally, we sum up all the terms and obtain
\begin{equation}
 \Omega=-\frac{1}{2}\sin\theta  \frac{E}{\sqrt{\frac{\Delta_{k_x}^2}{4} + \frac{\hbar^2k_y^2}{2m}\Delta_{k_x}}}
 \frac{1}{1+\frac{\frac{\Delta_{k_x}^2}{4} + \frac{\hbar^2k_y^2}{2m}\Delta_{k_x}}{\left(\frac{\hbar^2k_y^2}{2m}\right)^2}}
 \frac{\sqrt{\frac{2\hbar^2}{m\Delta_{k_x}}}}{1+\frac{2\hbar^2k_y^2}{m\Delta_{k_x}}}
 \left[\frac{\frac{\hbar^2 k_x}{m}}{\frac{\hbar^2k_y^2}{2m}} -  \frac{\hbar^2k_xk_y}{m\Delta_{k_x}}
 \frac{\frac{\Delta_{k_x}}{2}\frac{\hbar^2 k_y}{m}}{\left(\frac{\hbar^2k_y^2}{2m}\right)^2}\right].
\end{equation}
The expression in the square brackets is zero. Hence, $\Omega=0$.

\end{widetext}

\end{document}